\documentclass[aps,noshowpacs,twocolumn]{revtex4}
\usepackage{amsfonts}
\usepackage{amsmath}
\usepackage{amssymb}
\usepackage{graphicx}
\usepackage{bm}
\usepackage{color}

\begin{document}

\title{Rotational motion of dimers of Janus particles}
\author{Arghya Majee}
\email{majee@is.mpg.de}
\affiliation{Max-Planck-Institut f\"ur Intelligente Systeme, Heisenbergstr.\ 3, 70569 Stuttgart, Germany}
\affiliation{IV. Institut f\"{u}r Theoretische Physik, Universit\"{a}t Stuttgart, Pfaffenwaldring 57, 70569 Stuttgart, Germany}

\begin{abstract}
We theoretically study the motion of a rigid dimer of self-propelling Janus particles.
In a simple kinetic approach without hydrodynamic interactions, the dimer moves on a helical trajectory 
and, at the same time, it rotates about its center of mass. Inclusion of the effects of mutual 
advection using superposition approximation does not alter the qualitative features of the motion but merely 
changes  the parameters of the trajectory and the angular velocity.
\end{abstract}

\maketitle

\section{Introduction}

Self-propelling Janus particles (JPs) have gained increasing attention in recent
years from both theoreticians and experimentalists as they show a promising
route towards the understanding of the motion of living microorganisms 
\cite{How10, Cat12, Dre05,Elg15}. On the other hand, possible applications range from 
drug delivery to autonomous micromachines \cite{Sun08, Lao08}. Various types
of microswimmers have been built in recent years which mainly rely on the
nonuniform surface properties of the particle. Due to their asymmetric surface 
properties, these particles are able to generate their own gradient within an otherwise homogeneous 
medium and propel in this self-generated gradient. One particularly widely 
studied system in this context is chemical reaction 
driven self-propellers \cite{Pax04, Pax06, How07, Gol05}.

Of late, heating of half metal coated Janus particles has emerged as a possible way to
achieve self-propulsion. The metal cap can absorb energy from laser irradiation \cite{Jia10, Ber15} or 
ac magnetic field \cite{Bar13} and convert it into heat. Asymmetric thermal response of the capped and uncapped hemispheres 
then drives the colloid via self-thermophoresis. For a rotationally symmetric single particle the resulting
motion is linear initially; at longer times enhanced diffusion takes place due to rotational Brownian 
motion. However, a system of twin Janus particles with the one being tethered to the glass surface 
has been observed to rotate under laser irradiation \cite{Jia10}. Several other rotationally asymmetric systems 
have also been reported to show rotational movements. For example, circular motion of L-shaped asymmetric 
microswimmers on the substrate of a thin film and near channel boundaries has been reported \cite{Kum13}. 
Very recently, stable rotation was observed for a dimer system of chemically active Janus particles \cite{Wit15}.

In this paper we consider a dimer of rigidly attached Janus particles, which is free to 
move and to rotate in three dimensions and applies best to the case of self-thermophoretic
particles. Since usually the thermal conductivities of the solvent and the colloids are very 
close to each other, temperature field due to one particle is hardly affected by the presence 
of the neighbor particle. This is usually not the case for a dimer of catalytic Janus particles
as in this case the solutes can not penetrate the other particle and the concentration gradient is
affected by the presence of the second particle \cite{Moo16}. In our system, the orientation of the metal 
caps with respect to the dimer axis is kept arbitrary and each JP is treated in terms of 
a squirmer model where the local slip velocity at each particle surface is approximated by the first two Fourier 
components in an expansion in terms of the Legendre polynomial basis; see Eq. (\ref{eq0}) below and 
the following discussion. As a first approach we use a simple model without 
hydrodynamic interactions. Then we retain mutual advection and forces in terms of a superposition 
approximation for the flow fields created by the two particles of the dimer. 

\section{Dimer motion}

We consider the self-propulsion of a dimer, that is, of two Janus particles (squirmers) which are rigidly attached
to each other. Their motion arises from an effective slip velocity $ u_{s}(\theta)$, 
which is generated by a concentration or temperature gradient and depends only on the
polar angle $\theta$ with respect to the symmetry axis. A single particle moves at a velocity
\begin{equation}
 \mathbf{u}_{0}=u_0 \mathbf{n},
\end{equation}
where both its absolute value $u_0$ and direction $\mathbf{n}$ are determined by the weighted surface 
average of the slip velocity \cite{And89}.

\begin{figure}[!t]
\centering{\includegraphics[width=1.0\linewidth]{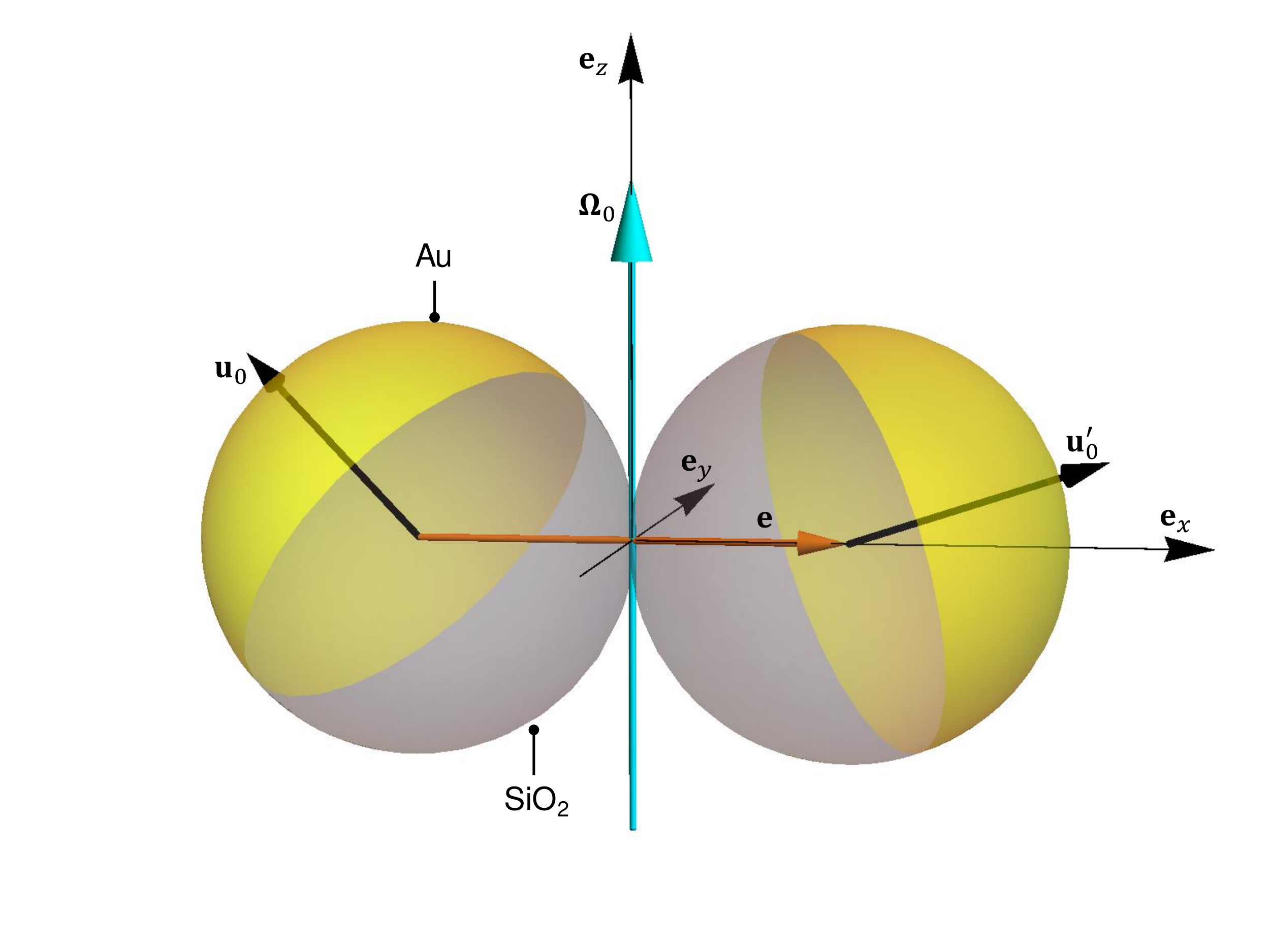}}
\caption{Sketch of a dimer which consists of two Janus particles attached to
each other in an arbitrary way. The co-ordinates are chosen in a way that
the angular velocity vector $\mathbf{\Omega}_0$ is directed along $z$-axis, the dimer axis $\mathbf{e}$ 
lies along $x$-axis, the relative velocity $\mathbf{u}_{0}-\mathbf{u}_{0}^{\prime}$ is in 
the $xy$-plane.}
\end{figure}

Two particles that are attached to each other, exert mutual forces and
result in a more complex motion which depends on the relative orientation of the particles with 
respect to the dimer axis. To describe the motion of this system, it is convenient to separate the 
center-of-mass velocity $\mathbf{U}$ and the relative motion of the JPs. For a rigid dimer, 
the relative motion reduces to its angular velocity $\mathbf{\Omega }$. 

We start with the simplest model which neglects advection. Then the center of mass motion is  
given by the mean value of the single-particle velocities, 
\begin{equation}
\mathbf{U}_{0}=\frac{\mathbf{u}_{0}+\mathbf{u}_{0}^{\prime}}{2}.  
\label{eq1}
\end{equation}%
The corresponding angular velocity,
\begin{equation}
\mathbf{\Omega }_{0}=\frac{\mathbf{u}_{0}-\mathbf{u}_{0}^{\prime}}{2a}\times 
\mathbf{e},
\label{eq2}
\end{equation}%
accounts for the orbital motion resulting from single-particle motion perpendicular to the dimer axis 
$\mathbf{e}$. As shown in Fig. 1, $\mathbf{e}$ points towards the primed particle.

\begin{figure}[!t]
\centering{\includegraphics[width=1.0\linewidth]{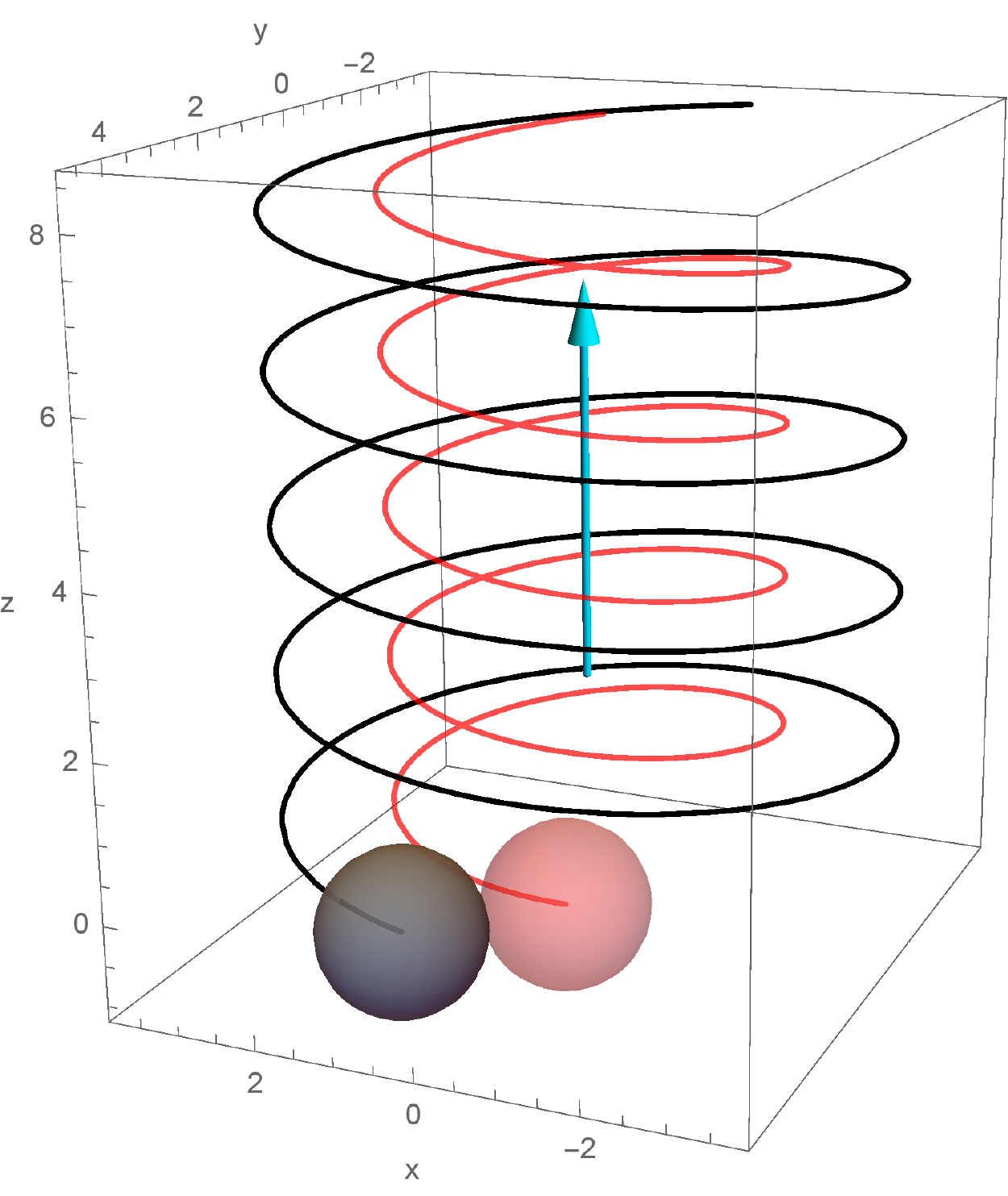}}
\caption{Sample trajectory for the centers of the two particles forming the dimer.
The arrow shows the direction of the angular velocity and the linear motion of the 
dimer. The particles rotate about the center of mass of the dimer at a rate similar 
to the rotation of the center of mass itself.}
\end{figure}

In the absence of additional torques and forces, the absolute values of both
linear and angular velocities are constant in time. The orientation of one
or the other, or of both, may change depending on the three vectors $\mathbf{%
u}_{0},\mathbf{u}_{0}^{\prime },\mathbf{e}$. The linear velocity $\mathbf{U}%
_{0}$ also rotates at an equal rate 
\begin{equation}
\frac{d}{dt}\mathbf{U}_{0}=\mathbf{\Omega }_{0}\times \mathbf{U}_{0}.
\label{eq3}
\end{equation}%
It turns out convenient to separate the velocity in two components $\mathbf{U}_{0}=\mathbf{U%
}_{0}^{\parallel }+\mathbf{U}_{0}^{\perp }$ which are parallel and perpendicular to the angular 
velocity. Then the trajectory of the center of mass
consists of a linear motion with velocity $\mathbf{U}_0^{\parallel }$, and a rotation in the plane perpendicular
to $\mathbf{\Omega }_0$; the latter is characterized by the angular
velocity $\Omega _{0}$ and a circular trajectory of radius  
\begin{equation}
R_{0}=U_0^{\perp }/\Omega _{0}.
\label{eq4}
\end{equation}
The dimer axis is always perpendicular to $\mathbf{\Omega }_{0}$ and thus
obeys the equation of motion 
\begin{equation}
\frac{d}{dt}\mathbf{e}=\mathbf{\Omega}_0\times \mathbf{e}.
\label{eq5}
\end{equation}%
Since the angular velocity $\mathbf{\Omega }_{0}$  is perpendicular on $\mathbf{e}$, it is constant in time. 

Thus the trajectories of the two JPs depend on the relative
orientation of $\mathbf{\Omega }_{0}$ and $\mathbf{U}_{0}$ and consist of
two contributions: The dimer shows translational motion at velocity $\mathbf{U}_0^{\parallel }$  along 
the vertical axis $\hat{\mathbf{z}}$ defined by $\mathbf{\Omega}_0=\Omega_0\hat{\mathbf{z}}$. 
Both $\mathbf{U}_0^\perp$ and $\mathbf{e}$
rotate about this axis at the angular velocity $\Omega_0$.  
The center of mass moves on a helical trajectory
\begin{equation}
\mathbf{R}_\text{cm} = U_0^\parallel t \hat{\mathbf{z}} + R_0 \hat{\mathbf{\varphi}}(\Omega_0 t) ,
\label{eq5a}
\end{equation}
where $\hat{\mathbf{\varphi}}$ is the local basis vector corresponding to the azimuthal angle $\varphi$. At the same time, 
the dimer axis $\mathbf{e}$ rotates in the plane perpendicular to $\hat{\mathbf{z}}$, such that each JP describes a 
helical trajectory, of radius $R_{\pm} = \sqrt{R_{0}^2 + a^2 \pm 2aR_{0}(\mathbf{e}\cdot\hat{\mathbf{\varphi}})}$. 

Typical trajectories of the two JPs are shown in Fig. 2. Note that the dimer axis $\mathbf{e}$ is constant with 
respect to the linear motion.  In the special case $\mathbf{u}_{0}=\mathbf{u}_{0}^{\prime}$, the angular
velocity vanishes and the dimer moves along a straight line, whereas $\mathbf{u}_{0}=-\mathbf{u}_{0}^{\prime}$ 
results in a simple rotation about the center of mass.

\section{Advection}

A moving particle gives rise to a characteristic velocity field in the surrounding fluid. In addition to its own motion, 
each particle of a dimer is advected in the velocity field $\mathbf{v}^{\prime }(\mathbf{r})$ of its neighbor and
is also subject to a mechanical force $\mathbf{F}$. Using Fax\'{e}n's law we find the linear velocity 

\begin{equation}
\mathbf{u}=\mathbf{u}_{0}+\frac{\mathbf{F}}{\xi }+\mathbf{v}^{\prime }(-2a%
\mathbf{e})+\frac{a^{2}}{6}\mathbf{\nabla }^{2}\mathbf{v}^{\prime }(-2a%
\mathbf{e}),
\label{eq11}
\end{equation}
where $\xi =6\pi \eta a$ is the usual Stokes friction factor with the viscosity $\eta$, and where 
$\mathbf{v}^{\prime }$ is evaluated at the distance 
$\mathbf{R}-\mathbf{R}^{\prime }=-2a\mathbf{e}$.

For the second particle we find the corresponding expression for $\mathbf{u}^{\prime }$ by exchanging 
primed and unprimed quantities, 
\begin{equation}
\mathbf{u}^{\prime }=\mathbf{u}_{0}^{\prime }+\frac{\mathbf{F}^{\prime }}{
\xi }+\mathbf{v}(2a\mathbf{e})+\frac{a^{2}}{6}\mathbf{\nabla }^{2}\mathbf{v}
(2a\mathbf{e}).
\label{eq12}
\end{equation}
Note the change of sign of the argument of the advection flow. For symmetry reasons, the forces cancel each other 
and are parallel  to the dimer axis, 
   \begin{equation}
   \mathbf{F} = F \mathbf{e}=-\mathbf{F}'.
   \label{eq12a}
   \end{equation}

\subsection{Fluid velocity field}

Due to some osmotic or catalytic effect, each particle induces an effective slip
velocity $u_{s}$ along its surface. For an axisymmetric Janus particle the slip velocity 
depends on the polar angle  $\theta$ only; the leading terms of an expansion in powers of $\cos\theta$ 
read as
  \begin{equation}
    u_{s}(\theta )=\frac{3}{2}u_0\sin \theta (1+\beta\cos\theta),
  \label{eq0}
  \end{equation} 
where the factor $\sin\theta$ is characteristic for a sphere \cite{Bla71}. The squirmer parameter $\beta$ is related 
to the long-range velocity field $\mathbf{v}\propto \beta  r^{-2}$ in the surrounding fluid, and to a large 
extent determines hydrodynamic interactions with neighbor particles or nearby solid boundaries; an active 
particle behaves like a ``puller'' for $\beta>0$ and like a ``pusher'' for $\beta<0$ \cite{Zot14, Llo10, Ish06}. 
The higher Fourier components in the expression for $u_s(\theta)$ are of the form $\beta_n dP_n(\cos \theta)/d\theta$ with $P_n$ 
being the Legendre polynomial of order $n$; the corresponding velocity field vanishes as $1/r^n$ and $1/r^{(n+2)}$ ($n>2$). 
However, these additional corrections would have little influence on the dimer motion since they would not alter the 
qualitative picture.

Throughout this paper we assume an axisymmetric slip velocity (\ref{eq0}) and we discard any single-particle angular 
motion. Eq. (\ref{eq0}) provides a boundary condition for the velocity field in the surrounding fluid,
\begin{align}
\mathbf{v}_T= &-\frac{1}{2}u_0\frac{a^{3}}{r^{3}}\left(1-3\hat{\mathbf{r}}\hat{\mathbf{r}}\right)\cdot\mathbf{n}\notag\\
              &-\frac{3}{2}\beta u_0\frac{a^2}{r^2}P_2(\mathbf{n}\cdot\hat{\mathbf{r}})\hat{\mathbf{r}}\notag\\
               &+\frac{3}{2}\beta u_0\frac{a^4}{r^4}\Big(P_2(\mathbf{n}\cdot\hat{\mathbf{r}})\hat{\mathbf{r}}
               -(\mathbf{n}\cdot\hat{\mathbf{r}})\left(1-\hat{\mathbf{r}}\hat{\mathbf{r}}\right)\cdot\mathbf{n}
               \Big),
\label{eq8}
\end{align}
where $\hat{\mathbf{r}}$ is the radial unit vector with respect to the particle center, $\mathbf{n}$ is the unit 
vector along the velocity vector $\mathbf{u}_0$, and $P_2$ represents the Legendre polynomial of second 
order \cite{Wur13}. Eq.~(\ref{eq8}) gives the usual flow field of a self-propelling particle; the first 
term on the right-hand side occurs for a particle with uniform surface properties in a constant driving field \cite{And89}, whereas the remainder accounts for the finite squirmer parameter.

In addition to $\mathbf{v}_T$, there is a velocity component arising from the force $\mathbf{F}$, 
\begin{equation}
\mathbf{v}_{F}=\left(\frac{3}{4}\frac{a}{r}\left(1+\hat{\mathbf{r}}\hat{\mathbf{r}}\right)+
               \frac{1}{4}\frac{a^{3}}{r^{3}}\left(1-3\hat{\mathbf{r}}\hat{\mathbf{r}}\right)\right)\cdot 
               \frac{\mathbf{F}}{\xi}.
\label{eq9}
\end{equation}
Note the presence of a long-range Stokeslet contribution proportional to $1/r $. The advection velocity in (\ref{eq12}) is given by the sum of the above
terms,%
\begin{equation}
\mathbf{v}=\mathbf{v}_{T}+\mathbf{v}_{F}.
\label{eq10}
\end{equation}
This velocity field is accompanied by the non-uniform pressure  
   \begin{equation}
    P = \frac{\mathbf{F}\cdot \mathbf{r}}  {4\pi r^{3}} -3\eta\beta u_0\frac{a^2}{r^3}P_2(\mathbf{n}\cdot\hat{\mathbf{r}}),
    \label{eq14a}
   \end{equation}
where the first term is related to the Stokeslet in $\mathbf{v}_{F}$ and the second one to the $r^{-2}$-contribution to the squirmer field. 

Similar expressions $\mathbf{v}'$ and $P'$ are obtained for the neighbor particle, by replacing the parameters $u_0$ and $\mathbf{F}$  
with corresponding primed quantities.

\subsection{Linear velocity}

The center-of-mass velocity of the dimer is given by 
\begin{equation}
\mathbf{U}=\frac{\mathbf{u}+\mathbf{u}^{\prime }}{2}.
\label{eq6}
\end{equation}
Inserting the single-particle velocities (\ref{eqA11}) and (\ref{eqA12}) we have 
\begin{align}
\mathbf{U}=&\left(1-\frac{\mathbf{Q}}{8}\right)\cdot\mathbf{U}_{0}\notag\\
            + & \frac{3\beta}{32}\Big(c' \mathbf{u}'_0  - c \mathbf{u}_0
              +  u_0 \frac{1-c^2}{2}\mathbf{e}  -    u'_0\frac{1-c'^2}{2}  \mathbf{e}   \Big),
\label{eq17}
\end{align}
where we have defined the quadrupole operator 
\begin{equation*}
\mathbf{Q}=1-3\mathbf{ee,}
\end{equation*}
and the orientation cosine
\begin{equation*}
  c  =\mathbf{n}\cdot\mathbf{e},\,\, \, \,  c'  =\mathbf{n'}\cdot\mathbf{e}.
\end{equation*}
Note that $\mathbf{U}$ depends on the relative orientation of the particle
axes with respect to the dimer axis $\mathbf{e}$.

\subsection{Angular velocity}

Similar to Eq.~(\ref{eq2}), in this case the angular velocity of the dimer with respect to
its center is given by 
\begin{equation}
\mathbf{\Omega }=\frac{\mathbf{u}-\mathbf{u}^{\prime }}{2a}\times \mathbf{e}.
\label{eq18}
\end{equation}
Inserting the single-particle velocities (\ref{eqA11}) and (\ref{eqA12}) we find
\begin{align}
\mathbf{\Omega} = \frac{17}{16}\mathbf{\Omega }_{0} 
               +\frac{3\beta}{32a}\Big( c' \mathbf{u}'_0 + c \mathbf{u}_0\Big)\times\mathbf{e}.
\label{eq18a}
\end{align}
In the first term we have used $\mathbf{Q}\cdot\mathbf{\Omega }_{0}=\mathbf{\Omega }_{0}$, which follows from 
the fact that $\mathbf{\Omega }_{0}$ is perpendicular to $\mathbf{e}$. The second term, which is proportional to the 
squirmer parameter $\beta$, results in a correction that is not parallel to $\mathbf{\Omega}_0$. 

The angular velocity $\mathbf{\Omega}$ is perpendicular on the dimer axis $\mathbf{e}$. According to the equation of motion  
\begin{equation}
\frac{d}{dt}\mathbf{e}=\mathbf{\Omega }\times \mathbf{e},
\label{eq7a}
\end{equation}
the dimer axis turns in the plane perpendicular to $\mathbf{\Omega}$ which, as a consequence, is constant in time, 
\begin{equation}
\frac{d}{dt}\mathbf{\Omega } = 0. 
\end{equation}

\subsection{Mutual forces}

So far the mutual forces (\ref{eq12a}) are not known; their strength $F$ is determined from the condition that the single-particle
velocities have the same component along the dimer axis, 
   \begin{equation}
   (\mathbf{u-u}' )  \cdot \mathbf{e}=0.
   \label{eq52}
   \end{equation}
Inserting the single-particle velocities $\mathbf{u}$ and $\mathbf{u}'$ given in the Appendix and solving for $F$, we find the force 
acting on the unprimed particle,
   \begin{equation}
   \frac{F}{\xi} =  \frac{7}{6}(\mathbf{u}'_0-\mathbf{u}_0 )  \cdot \mathbf{e} 
                           - \frac{\beta}{4}\left(u_0 P_2(c)+u_0' P_2(c')  \right) .
   \label{eq54}
   \end{equation}

\section{Discussion}

In the absence of advection, the dimer moves on a helical trajectory (\ref{eq5a}), as illustrated in Fig. 1. 
This picture remains valid when including advection corrections, albeit with numerically modified linear and 
angular velocities $\mathbf{U}$ and $\mathbf{\Omega}$ as expressed by Eqs.~(\ref{eq17}) and (\ref{eq18a}). 

The helical trajectory is conserved because of the symmetry of the advection field. More complex trajectories would arise if the 
angular velocity had a component proportional to $\mathbf{e}$, in other words, if the dimer rotated about its axis. Then 
$\mathbf{\Omega}$ is no longer a constant, resulting in a more intricate motion.

Here we mention two effects that would result in a time dependent angular velocity. First, real Janus particles are not perfectly 
axisymmetric. In general, their slip velocity comprises a constant in azimuthal direction, resulting in a rotation about the particle 
axis $\mathbf{n}$ with angular velocity $\mathbf{\omega}_0$. Then the angular velocity of the dimer comprises a term proportional 
to its axis, $\Omega_\parallel =(\mathbf{\omega}_0+\mathbf{\omega}'_0)\cdot\mathbf{e}$, which leads to a more complex trajectory. 

Second, in this paper we have only addressed mutual advection of the two Janus particles, and neglected the influence of their 
activity. In the case of self-thermophoresis, for example, the slip velocity on the surface of one particle depends not only on its own 
temperature gradient but also on that of the neighbor, 
   \begin{equation}
   \mathbf{u}_s = \mu(\hat{\mathbf{r}})  (\nabla_\parallel T +\nabla_\parallel T'),
   \label{eq70}
   \end{equation}
where $\nabla_\parallel$ is the gradient component parallel to the particle surface. Since in general the mobility $\mu$ takes different 
values on the two hemispheres of a Janus particle, the term $\nabla_\parallel T'$ induces an angular velocity $\omega_0$ about 
the dimer axis. If the two particles are heated at different temperatures, or show different surface properties, their contributions to 
$\Omega_\parallel$ do not cancel, and there is a net rotational motion of the dimer about its axis $\mathbf{e}$. 

For usual Janus particles these effects are small, and probably could not be distinguished from the rotational diffusion. Thus we are 
led to the conclusion that advection is the dominant interaction and that dimers in bulk solution move on helical trajectories.  One 
should also keep in mind that gravity can play a role \cite{Cam13}. Because of the heavy metal cap on one hemisphere, the center of mass 
does not coincide with the geometric center. Then gravity exerts a torque on the dimer, which in particular results in a time-dependent 
angular frequency vector $\Omega$.

We use the superposition approximation, in other words we neglect higher reflections between the two particles.
The reflection method in principle is valid for large distances.  But as discussed before, the temperature field
is rather insensitive to the presence of a neighbor particle due to a small conductivity difference between 
the particle and the solvent. On the contrary, the hydrodynamic flow field is sensitive to the presence of 
the second particle. However, it becomes more important when the particles forming the dimer move with respect to 
each other. For a rigidly attached dimer, the near-field effects are expected to be less significant. Therefore, a helical 
like trajectory appears to be a generic feature for such rigid systems. Improving either the superposition approximation by 
considering higher reflections or considering higher order terms in Fax\'{e}n's law would render quantitative changes. 
But the relatively small linear corrections in Eqs.~(\ref{eq17}) and (\ref{eq18a}) (note the prefactor $3/32$ in the terms 
proportional to $\beta$) suggest that higher order terms are even smaller and do not change the 
qualitative picture. Similar numerical changes would result when improving the superposition approximations 
by applying the method of reflections. Thus, the overall trajectory of a rigid dimer is expected to be close to the helical one
unless the additional effects mentioned in the preceding paragraphs are large enough to alter the picture.

We conclude with a remark on the forces (\ref{eq12a}) exerted by one Janus particle on the other. These mutual forces cancel 
in the linear and angular velocity, and thus are of little relevance for a rigid dimer. They are important, however, for 
particles linked by flexible DNA strands, where the length of the macromolecular bridge is determined by equilibrating 
the entropic force with the mutual force between two JPs, or between a JP and a passive colloid; such hybrid systems 
have been designed recently \cite{Sch15}. Although, one should keep in mind that for a flexible coupling, the two 
particles also exert torques on each other which results in a more complex rotation than that of a rigid dimer.

\begin{acknowledgments}
Stimulating discussions with Alois W\"{u}rger are gratefully acknowledged.
\end{acknowledgments}

\section*{Author contribution statement}
A. M. conceived the problem, performed all the calculations, and wrote the manuscript.

\appendix

\section{Single-particle velocities}

Here we give the advection corrections to the velocities $\mathbf{u}$ and $\mathbf{u}'$
of the Janus particles forming the dimer. We evaluate the  different contributions to the right hand 
sides of Eqs.~(\ref{eq11}) and (\ref{eq12}). 

We start with the velocity field $\mathbf{v}$ created by the 
unprimed particle at the position of its neighbor (i.e., $\mathbf{r}=2a\mathbf{e}$). There are two 
contributions, $\mathbf{v}(2a\mathbf{e})=\mathbf{v}_T(2a\mathbf{e})+\mathbf{v}_F(2a\mathbf{e})$, which 
arise from the particle's self-propulsion and the force exerted by its neighbor, respectively.  
From the explicit expressions given in Eqs.~(\ref{eq8}) and (\ref{eq9}), one readily obtains the advection terms 
\begin{align}
\mathbf{v}_T(2a\mathbf{e})=&-\frac{3}{8}\beta u_0P_2(c)\mathbf{e}-\frac{\mathbf{Q}\cdot\mathbf{u}_0}{16}\notag\\
                           &+\frac{3\beta u_0}{32}\left[P_2(c)\mathbf{e}-c(1-\mathbf{ee})\cdot\mathbf{n}\right],
\label{eqA2}
\end{align}
and 
\begin{equation}
\mathbf{v}_F(2a\mathbf{e})=\left( \frac{3\left( 1+\mathbf{ee}\right) }{8}+%
\frac{\mathbf{Q}}{32}\right) \cdot \frac{\mathbf{F}}{\xi}.
\label{eqA1}
\end{equation}
The correction term is evaluated by using Stokes' equation $\eta\mathbf{\nabla }^{2}\mathbf{v}=\mathbf{\nabla }P$ with 
the expression for the pressure given in Eq.~(\ref{eq14a}). Thus we find the contribution arising from self-propulsion,
   \begin{equation}
   \mathbf{\nabla}^{2}\mathbf{v}_{T}=9\beta u_0\frac{a^2}{r^4}\Big(P_2(\mathbf{n}\cdot\hat{\mathbf{r}})\hat{\mathbf{r}}
   -(\mathbf{n}\cdot\hat{\mathbf{r}})(1-\hat{\mathbf{r}}\hat{\mathbf{r}})\cdot\mathbf{n}\Big),
   \label{eqA3}
   \end{equation}
and similarly that due to the mutual force,
\begin{equation}
  \mathbf{\nabla }^{2}\mathbf{v}_{F}=(1-3\hat{\mathbf{r}}\hat{\mathbf{r}})\cdot\frac{\mathbf{F}}{4\pi\eta r^3}.
\label{eqA4}
\end{equation}
Putting $\mathbf{r}=2a\mathbf{e}$ and using the definitions for $c$ and $\textbf{Q}$, we have
   \begin{equation}
   \frac{a^2}{6}\mathbf{\nabla}^{2}\mathbf{v}_{T}(2a\mathbf{e})=\frac{3\beta u_0}{32}\Big(P_2(c)\mathbf{e}
   -c(1-\mathbf{ee})\cdot\mathbf{n}\Big),
   \label{eqA5}
   \end{equation}
and
\begin{equation}
  \frac{a^2}{6}\mathbf{\nabla }^{2}\mathbf{v}_{F}(2a\mathbf{e})=\frac{\mathbf{Q}\cdot\mathbf{F}}{32\xi}.
\label{eqA6}
\end{equation}

We give the corresponding quantities for the primed particle at the relative position $\mathbf{r}=-2a\mathbf{e}$ (note that 
the unit vector $\mathbf{e}$ points from the unprimed to the primed particle):
\begin{align}
\mathbf{v'}_T(-2a\mathbf{e})=&\frac{3}{8}\beta u_0'P_2(c')\mathbf{e}-\frac{\mathbf{Q}\cdot\mathbf{u'}_0}{16}\notag\\
                           &+\frac{3\beta u_0}{32}\left[-P_2(c')\mathbf{e}+c'(1-\mathbf{ee})\cdot\mathbf{n'}\right],
\label{eqA8}
\end{align}
\begin{equation}
\mathbf{v'}_F(-2a\mathbf{e})=\left( \frac{3\left( 1+\mathbf{ee}\right) }{8}+%
\frac{\mathbf{Q}}{32}\right) \cdot \frac{\mathbf{F'}}{\xi },
\label{eqA7}
\end{equation}
   \begin{equation}
   \frac{a^2}{6}\mathbf{\nabla}^{2}\mathbf{v'}_{T}(-2a\mathbf{e})=\frac{3\beta u'_0}{32}\Big(-P_2(c')\mathbf{e}
   +c'(1-\mathbf{ee})\cdot\mathbf{n'}\Big),
   \label{eqA9}
   \end{equation}
   and,
\begin{equation}
  \frac{a^2}{6}\mathbf{\nabla }^{2}\mathbf{v'}_{F}(-2a\mathbf{e})=\frac{\mathbf{Q}\cdot\mathbf{F'}}{32\xi}.
\label{eqA10}
\end{equation}

Finally, inserting Eqs.~(\ref{eqA2}-\ref{eqA10}) in (\ref{eq11}) and (\ref{eq12}), one obtains the single-particle velocities 
\begin{align}
\mathbf{u}=&\mathbf{u}_0 - \frac{\mathbf{Q}\cdot\mathbf{u'}_0}{16}
                         + \frac{\mathbf{F}}{\xi}    +\frac{5}{8} \frac{\mathbf{F'}}{\xi}            \notag\\
                   &  +\frac{3\beta}{16} \Big( u'_0P_2(c')\mathbf{e} + c'(1-\mathbf{ee})\cdot\mathbf{u'}_0  \Big),
\label{eqA11}
\end{align}
and
\begin{align}
\mathbf{u'}=&\mathbf{u'}_0 - \frac{\mathbf{Q}\cdot\mathbf{u}_0}{16} 
                                  + \frac{\mathbf{F'}}{\xi} +\frac{5}{8} \frac{\mathbf{F}}{\xi}     \notag\\
           &-\frac{3\beta}{16}   \Big( u_0P_2(c)\mathbf{e} + c(1-\mathbf{ee})\cdot\mathbf{u}_0  \Big).
\label{eqA12}
\end{align}
Using $\mathbf{F'}=-\mathbf{F}$  further simplifies  the force terms.

\end{document}